\documentclass{ws-p10x7} 

\begin{document} 
 
\title{Review on B$^0$- $\overline{\rm \bf{B^0}}$ mixing and b-lifetimes measurements at CDF/LEP/SLD} 
 
\author{A. Stocchi} 
 
\address{Laboratoire de l'Acc\'el\'erateur Lin\'eaire 
IN2P3-CNRS et Universit\'e de Paris-Sud,\linebreak BP 34, F-91898 Orsay Cedex, 
(stocchi@lal.in2p3.fr)} 
 
 
 
\twocolumn[\maketitle\abstract{ A review of the results on B$^0$- $\overline {\rm B^0}$ mixing and b-lifetimes  
obtained by CDF, LEP and SLD collaborations is presented with special emphasis on B$^0_s$- $\overline {{\rm B}^0_s}$ mixing.}] 
 
\section{Introduction}
\vspace{-0.2cm} 
In the last decade, new weakly decaying B-hadrons have been observed (B$^0_s, {\rm B}_c, \Lambda^0_b, \Xi_b$) and their production and decay properties have been intensively studied. In this context CDF (operating at TeVatron), ALEPH, DELPHI, L3, OPAL (operating at LEP) and SLD (operating at SLC) experiments have played a central role. This has been made possible owing to the excellent performance both of the machines and of the detectors. Above all, these measurements would have not been possible without the development of Silicon detectors. 
\vspace{-0.3 cm} 
\section{B hadron lifetimes} 
The measurement of the lifetimes of the different B hadrons is an important test of the B decay dynamics\cite{ref1}. 
\begin{table*}[t] 
\centering 
\caption{Lifetime ratios results \label{tab:un}} 
\begin{tabular}{|l|l|l|l|p{50pt}|}  
\cline{1-3} 
\hline 
\raisebox{0pt}[8pt][6pt]{}   
Lifetime ratios & 
Osaka 2000 & 
Tampere 1999&Theory\\[3pt] 
\hline  
\raisebox{0pt}[8pt][6pt]{$\tau (B^-)/\tau (B^0)$} 	& $1.070 \pm 0.020$ 	& $1.065 \pm 0.023$ & 1.0 - 1.1\\[6pt]  
\raisebox{0pt}[8pt][6pt]{$\tau (B^0_s)/\tau (B^0)$} &$0.945 \pm 0.039$ & 
$0.937 \pm 0.040$ & 0.99 - 1.01\\[6pt] 
\raisebox{0pt}[8pt][6pt]{$\tau (b - bary)/\tau (B^0)$}& $0.780 \pm 0.035$ & 
$0.773 \pm 0.036 $ & 0.9 - 1.0\\[6pt] \hline 
\end{tabular} 
\end{table*}


The main improvements, since last year, have been obtained in the determination of B$^0_d$ and B$^+$ lifetimes. The results\cite{refA} are given in Table 1. In future no improvements are really expected before the start of the new phase at TeVatron. Few conclusions can be drawn. The charged B-mesons live longer than the neutrals. This effect is now established at 3.5 $\sigma$. The lifetimes of the two neutral B mesons (B$^0_d, {\rm B}^0_s$) are equal within $1\sigma$. To observe any possible (and unexpected difference)  
new data are needed. The b-baryon lifetime problem still exists (since 3-4 years). b-baryons live shorter than b-mesons,  
as expected, but the magnitude of the effect is now more than  3 $\sigma$ away from the lower edge of  
the predictions\cite{ref1}. This  should push for a better understanding of the theo\-ry independently  
of a possible improvement of the experimental accuracy. 
\vspace{-0.3cm} 
\section{Lifetime Difference: $\Delta\Gamma_s$} 
The ratio between the differences of the widths and of the masses of the B$^0_s$-$\overline {\rm{B}}^0_s$  
system mass eigenstates is (naively): $\Delta \Gamma_s/\Delta m_s \sim 3/2 \pi (m_b/m_t)^2$. If $\Delta m_s$ is too large (and so, difficult to measure), $\Delta \Gamma_s$ can eventually gives access to it. Unfortunately the theoreti\-cal error attached to the evaluation of $\Delta \Gamma_s$ is still quite large, of the order of 50 \%. Recent theoretical calculations predict $ \Delta \Gamma _s/\Gamma_s$ in the range (5-10) \%\cite{ref2}.\\ 
From the experimental point of view, the combination of LEP and CDF  results gives\cite{refB} (assuming $\tau (B^0_d) = \tau (B^0_s)$): 
$$\Delta \Gamma_s/\Gamma_s = 0.16^{+0.16}_{-0.13} ; <  0.31 \ {\rm at} \ 
 95 \% \ {\rm C. L.}$$ 

\section{B$^0 - \overline{\rm B^0}$ mixing : $\Delta m_d, \Delta m_s$} 
In the Standard Model, B$^0_q - \overline{\rm B}^0_q$ ($q = d,s$) mixing can be expressed by the following formula: 
\begin{eqnarray} 
\Delta m_q &=& G^2_F/6\pi^2 m^2_W \eta_c S (m^2_t/m^2_W) \vert V_{tq} \vert^2 
 \nonumber\\ 
&&m _{Bq} f^2_{Bq} B_{Bq}  
\end{eqnarray} 
where S$(m^2_t/m^2_W)$ is the Inami-Lim function, $m_t$ is the $\overline{MS}$ top mass and $\eta_c$ is a QCD correction factor obtained at NLO order in perturbative QCD. The measurement of $\Delta m_d, (\Delta m_s)$ gives access to $V_{td}, (V_{ts})$ CKM matrix elements and thus to the $\overline \rho$ and $\overline \eta$ parameters\footnote{$\bar \rho$ and $\bar \eta$ are related to the original $\rho$ and $\eta$ parameters 
$\bar \rho = \rho (1 - \lambda^2 /2), \bar \eta = \eta(1 - \lambda^2 /2)$.}  
of the Wolfenstein parametrization. We can write: 
\vspace{-1mm} 
\begin{eqnarray} 
\begin{array}{ll} 
\Delta m_d \sim  A^2 \lambda^6 [(1-{\overline \rho})^2 + {\overline \eta }^2)] f^2_{Bd} B_{Bd}  \\ 
\Delta m_s \sim   A^2 \lambda^4 f^2_{Bs} B_{Bs}  \\ 
\rightarrow \Delta m_s \sim  1/ \lambda^2 \Delta m_d \sim  20 \Delta m_d  \\ 
\Delta m_d/\Delta m_s \sim  \lambda^2/\xi^2 [(1-\overline \rho)^2 +\overline  
\eta^2] \\ 
{\rm with}\  \xi = \frac{f_{Bs}\sqrt{B_{Bs}}}   {f_{Bd} \sqrt{B_{Bd}}} 
\end{array} 
\end{eqnarray} 
 
The interest of measuring both $\Delta m_d$ and $\Delta m_s$ comes from the fact that the ratio $\xi$ is better determined from theory than the individual quantities entering into its expression.

The analyses presented here measure $\Delta m_q$ by looking at the time dependence behaviour of the oscillations : 
\vspace{-2mm} 
\begin{equation} 
P (B^0_q \rightarrow \stackrel{(-)}{B^0_q}) = \frac {1}{2 \tau} e^{-t/\tau} 
(1 (\pm) cos \Delta m_q t) 
\end{equation} 
\vspace{-1cm} 
\subsection{$\Delta m_d$ results} 
The time variation of the B$_d$ oscillation has been observed for the first time at LEP. In the last years the precision has impressively improved  down to  
3 \% giving\cite{ref6} : 
$$\Delta m_d = 0.486 \pm 0.015 \ {\rm ps}^{-1}  \ \ LEP/SLD/CDF$$ 
\noindent 
which is an average of 26 measurements ! 
 
The recent CLEO $\chi_d$ measurement is in agreement with this value giving\cite{ref6} the final result of : $\Delta m_d = 0.487 \pm 0.014 {\rm ps}^{-1}$. Improvements on this result are expected in the coming years from B-factories. 
\vspace{-0.1cm} 
\subsection{$\Delta m_s$ results} 
Since ${\rm B}_s$ mesons are expected to oscillate  20 times faster than ${\rm B}_d$, it is fundamental to have the best possible resolution on the proper time reconstruction. For details on the analyses see the contributions of P. Coyle\cite{ref3} and T. Usher\cite{ref3} in these proceedings. No experiment has observed an oscillation signal. A procedure has been set to combine the different analyses and to get a limit or eventually to quantify the evidence for a ``combined'' signal. This is done in the framework of the amplitude method\cite{ref5} which consists in modifying the last part of eq.(3) using: 1 $\pm $A cos $\Delta m_s t$. At any given value of $\Delta m_s$, A and $\sigma_A$ are measured. ${\rm A}=1$ and not compatible with ${\rm A}=0$ indicates an oscillation  signal at the corresponding value of $\Delta m_s$. The values of $\Delta m_s$ excluded at 95 \% C. L. are those satisfying $A(\Delta m_s) + 1.645 \sigma_A (\Delta m_s) < 1$. It is also possible to define the sensitivity as the value of $\Delta m_s$ corresponding to 1.645 $\sigma_A (\Delta m_s)$ = 1. The main actors in this work are LEP and SLD collaborations.

Figure 1 shows the evolution of the combined sensitivity which has dramatically improved during the years. Figure 2 gives the combined plot of the amplitude values as a function of $\Delta m_s$\cite{ref6}. The results are: 
\begin{center} 
$\Delta m_s > 14.9\  \rm{ps}^{-1}{\rm at\  95\% \ C.L}$ \\ 
{sensitivity at 17.9 ps$^{-1}$}. 
\end{center} 
 
 
\begin{figure}[h] 
\vspace{-0.7cm} 
\epsfxsize150pt 
\figurebox{}{}{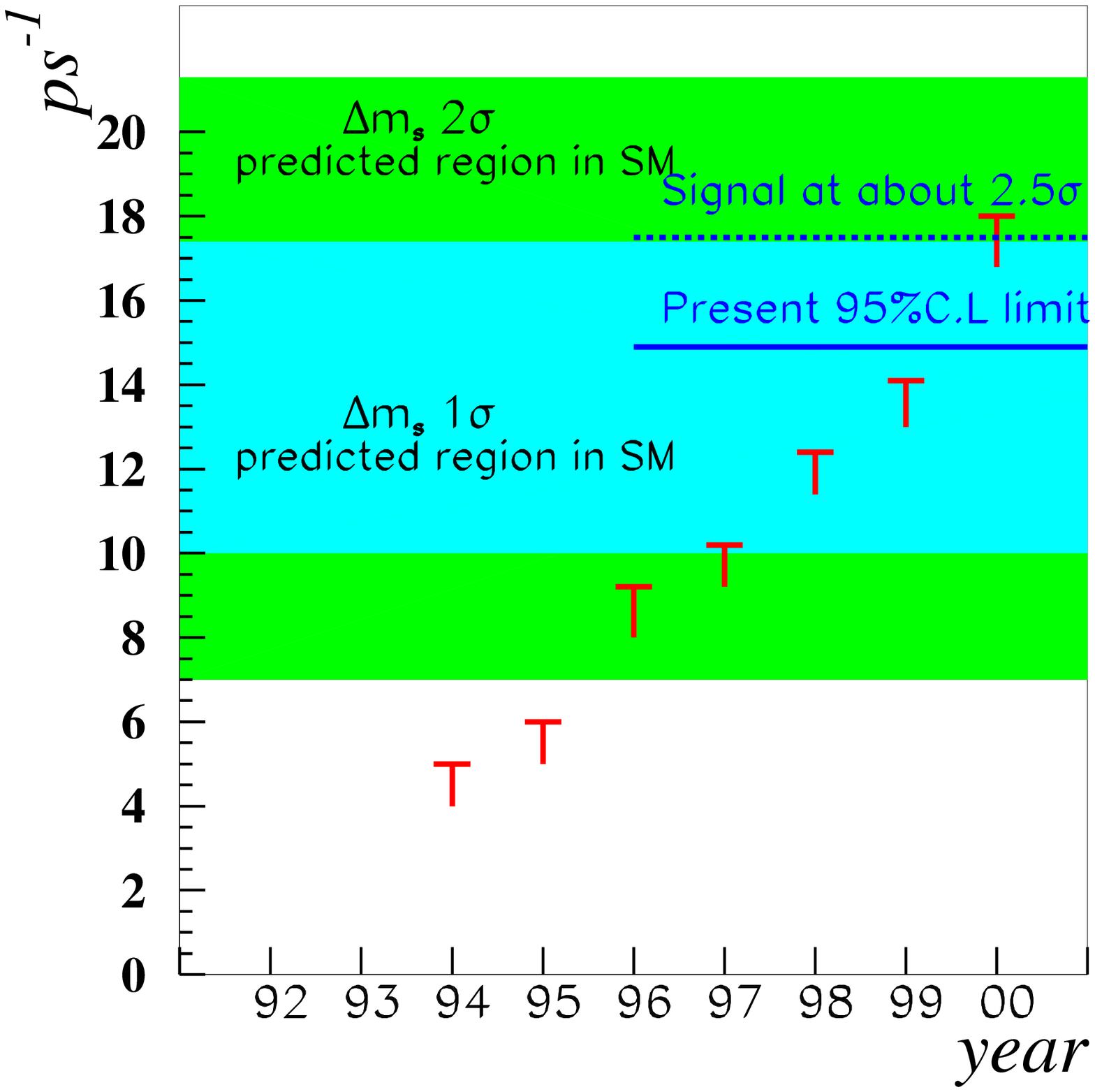} 
\caption{The evolution of the combined $\Delta m_s$ sensitivity over the years.} 
\label{fig:1} 
\end{figure} 
 
A ``signal'' bump is visible at around  
$\Delta m_s = 17.7 $ ps$^{-1}$ with a significance at 2.5 $\sigma$ level. The probability of a background fluctuation greater of equal to the one observed, and at any $\Delta m_s$ value, has been evaluated to be about 2.5 \%. 
This result is still expected to improve during next months by continuing the progress in LEP/SLD analyses. 
 
The impact of this result on the determination of the unitarity triangle parameters is shown in Figure 3. Using the constraint  
coming from the measurements of V$_{\rm ub}$, $\Delta{\rm m_d}$, $ \mid \epsilon_K \mid $ and $\Delta {\rm m_s}$ we obtain\cite{ref7}: 
 
\vspace{-6mm} 
\begin{eqnarray*} 
\bar \rho   &=& 0.206 \pm 0.043 ;  \bar \eta =  0.339 \pm  0.044 \nonumber \\ 
sin 2 \beta &=& 0.723 \pm  0.069 ;  sin2\alpha = -0.28  \pm  0.27 \nonumber \\ 
&{\rm and}& \gamma = (58.5 \pm 6.9)^{\circ} 
\end{eqnarray*} 
\vspace{-1cm} 
\section*{Conclusions} 
\vspace{-0.3cm} 
The different B-lifetimes have been measured at the few percent level  
($\sim$ 1.5 \% for B$^+$ and B$^0_d$ and $\sim 4 \% $ for B$^0_s$ and $\Lambda^0_b$). A clear experi\-mental hierarchy has been  
established. The frequency of the B$^0_d$ meson oscillation ($\Delta m_d$) has been measured with a 3 \% precision. As far as the  
B$_s$ oscillation is concerned the combined sensitivity is now at 17.9 ps $^{-1}$ and a possible signal with a 2.5 $\sigma$  
significance at about $\Delta m_s = 17.7$ ps$^{-1}$ has been observed. This result is still expected to be improved  
during the coming months. Let's wait a bit for claiming the observation of B$^0_s-\overline{\rm B^0_s}$ oscillations ! 
 
\begin{figure} 
\vspace{-0.8cm} 
\epsfxsize163pt 
\figurebox{}{}{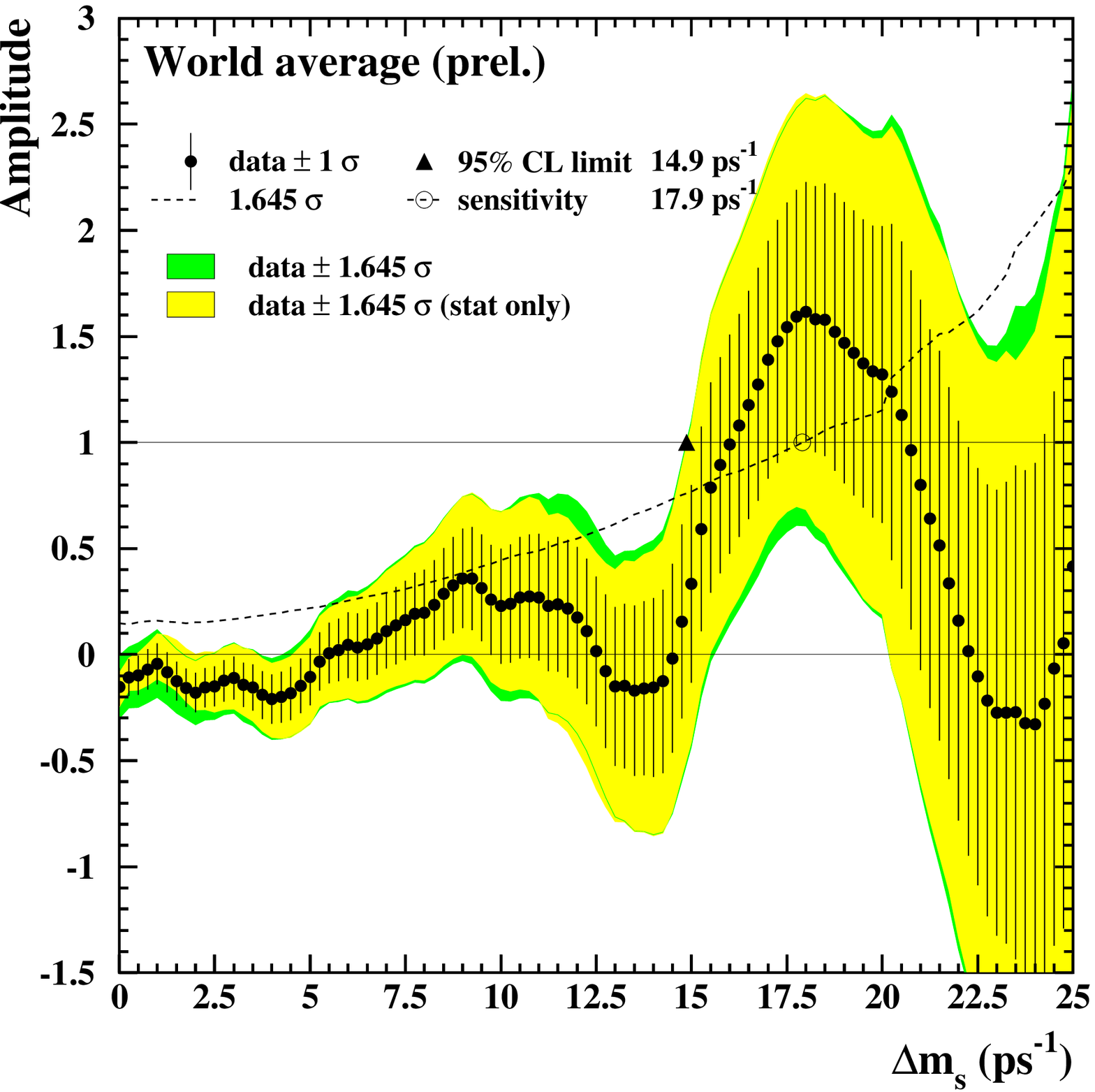} 
\caption{Variation of the combined amplitude versus $\Delta m_s$. The points with error  
bars are the data ; the lines show the 95 \% C. L. curves (in dark when systematics are included). The dotted curve indicates the sensitivity.} 
\label{fig:2} 
\end{figure} 
 
\begin{figure}[h] 
\vspace{-0.5cm} 
\epsfxsize180pt 
\figurebox{}{}{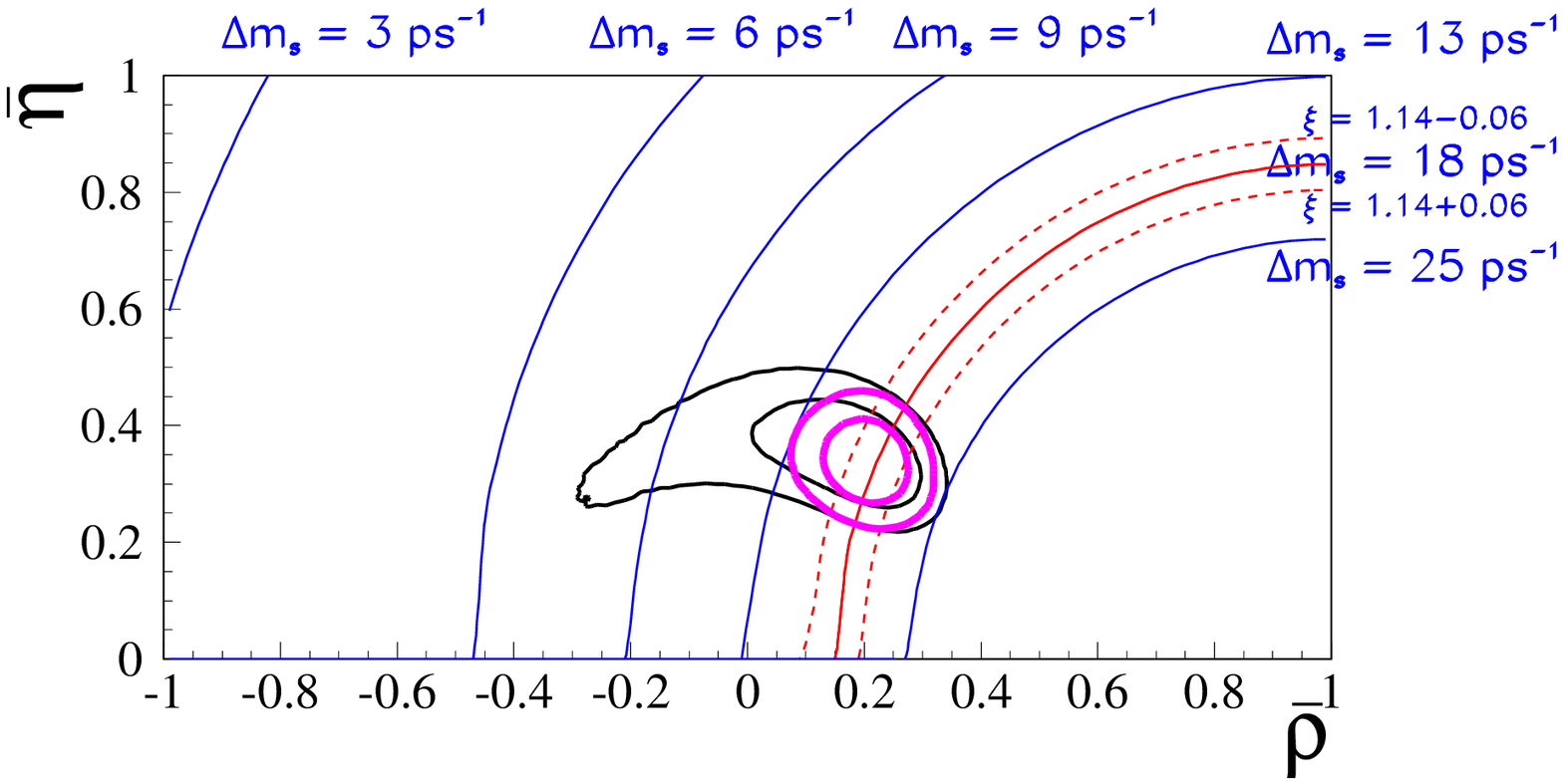} 
\caption{The allowed regions for $\overline \rho$ and  $\overline \eta$ using the constraints given by the measurements  
of $\vert \epsilon_k \vert, \vert V_{\rm ub} \vert /\vert  V_{\rm cb}\vert$ and  
$\Delta {\rm m_d}$ at 68 \% and 95 \% probability are shown by the thin contour lines.  
The constraint due to $\Delta {\rm m_s}$ is not included. Selected regions for   
$\overline \rho$ and  $\overline \eta$ when the constraint due to $\Delta {\rm m_s}$  
is included have been superimposed using thick lines.} 
\label{fig:3} 
\end{figure} 
\vspace{-0.4cm} 
\section*{Acknowledgement} 
\vspace{-0.2cm} 
Congratulations to all the collegues of CDF/LEP/SLD which have made all of it possible !  
Thanks to P. Roudeau for the careful reading of the manuscript.

\end{document}